\documentclass{PoS}
\title{New insights into soft gluons and gravitons}

\ShortTitle{New insights into soft gluons and gravitons}

\author{\speaker{Chris D. White}\\
        University of Glasgow\\
        E-mail: \email{Christopher.White@glasgow.ac.uk}}

\abstract{The study of gluon radiation in QCD, in the limit of small (``soft'')
momentum, remains an active research area, with a variety of phenomenological 
and theoretical applications. Soft gluon emission leads to large logarithms in 
perturbation theory which have to be summed up to all orders in the coupling, 
and also governs the structure of infrared singularities. Recently, new 
techniques and mathematical structures have been discovered, which enhance our 
understanding of these all-order properties. This contribution will review a 
number of key topics, including the relationship between QCD and gravity.}

\FullConference{36th International Conference on High Energy Physics,\\
		July 4-11, 2012\\
		Melbourne, Australia}

\begin{document}

\section{Introduction}
\begin{figure}
\begin{center}
\scalebox{0.8}{\includegraphics{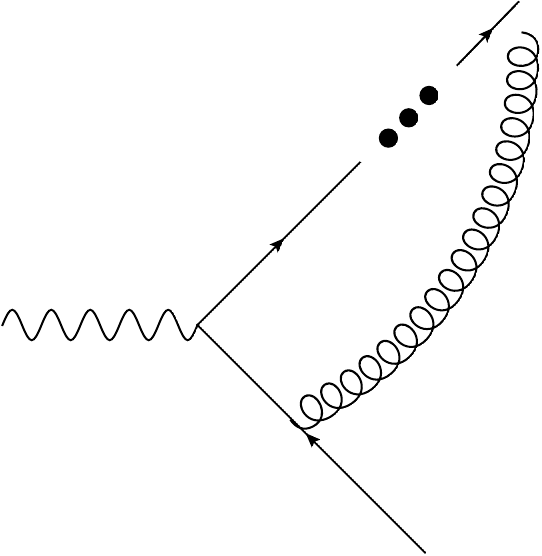}}
\caption{Example interaction, in which soft gluons may be emitted.}
\label{fig:pair}
\end{center}
\end{figure}

It is well-known that scattering amplitudes in quantum field theory are beset 
by infrared divergences. Consider, for example, the interaction shown in
figure~\ref{fig:pair}, in which a vector boson splits into a quark pair.
Either the final state quark or anti-quark may emit gluon radiation, and the 
Feynman rules in position space tell us that we must integrate over all 
positions of the emitted gluon. This integral goes out to infinity, and 
one may show from the behaviour of the integrand in four space-time dimensions
that the region at infinity is associated with a (long-distance) divergence.
We may think of such gluons as having an infinite Compton wavelength, and 
by the uncertainty principle this corresponds to the emission of gluons with
zero momentum. Thus, these divergences are usually referred to as {\it infrared
(IR) divergences}, and the gluons themselves as {\it soft}, to distinguish
them from the {\it hard} emitting particles that emerge from the interaction.
Such gluons may be real or virtual, and the above remarks make clear that 
IR divergences are a general feature of quantum field theories, including
QED, QCD and quantum gravity.\\

Infrared singularities are important for a number of theoretical and 
phenomenological reasons. They are related to the structure of large 
logarithms in perturbation theory, which must be summed up to all orders in
order to obtain sensible results for many collider observables 
(see~\cite{Sterman:1986aj,Catani:1989ne,Korchemsky:1993uz,Becher:2006nr,
Laenen:2008gt} for a number of different approaches). Furthermore, there
are a number of unproven conjectures regarding infrared divergences, such as
the so-called {\it dipole formula} in QCD~\cite{Becher:2009cu}, 
which we will see in what follows. It has by now been well-established that 
scattering amplitudes factorise, such that they have the schematic 
form~\cite{Collins:1980ih}
\begin{equation}
A=H\cdot S\cdot \frac{\prod_i J_i}{\prod_i{\cal J}_i}.
\label{ampfac}
\end{equation}
Here $H$ is the {\it hard interaction}, and is infrared finite; $S$ is the
{\it soft function}, which collects all soft singularities; $J_i$ is a 
{\it jet function}, which collects hard collinear singularities associated
with outgoing particle $i$. Finally, ${\cal J}_i$ is an {\it eikonal jet 
function}, which corrects for the double-counting of soft-collinear 
singularities associated with particle $i$~\footnote{In quantum gravity, jet
functions are not present, as collinear singularities cancel upon combining
all diagrams~\cite{Weinberg:1965nx,Akhoury:2011kq}.}. Furthermore, it is known
in a variety of theories that the soft function exponentiates, where the 
exponent may itself be given a Feynman diagram interpretation. That is, one
may write
\begin{equation}
S=\exp\left[\sum_W W\right],
\label{Sexp}
\end{equation}
where $\{W\}$ is a set of special diagrams known as {\it webs}~\footnote{This 
name was first introduced in the context of two-parton scattering in 
QCD~\cite{Gatheral:1983cz}, but we here adopt the more general terminology 
of~\cite{Gardi:2010rn}.}. Their nature depends upon the theory. \\

\section{Webs in QED and QCD}
\begin{figure}
\begin{center}
\scalebox{0.8}{\includegraphics{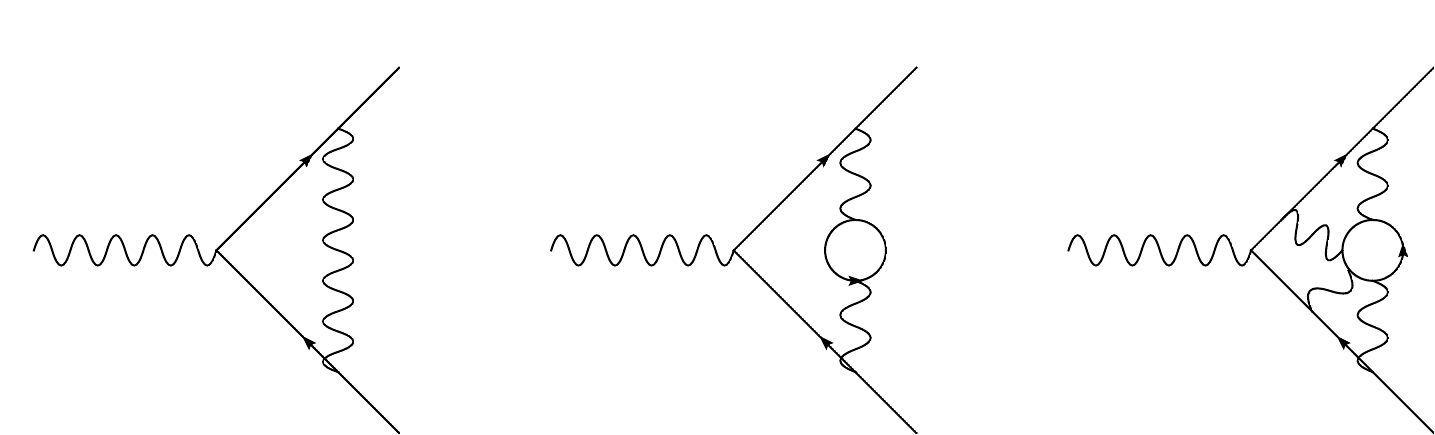}}
\caption{The first few webs in QED. Upon removing the external lines, one is
left with connected subdiagrams.}
\label{fig:QED}
\end{center}
\end{figure}

In QED, one may show that the exponent of the soft function contains only
connected subdiagrams. By a subdiagram, we mean the graph that remains when
the hard external lines have been removed. Examples of QED ``webs'' are shown
in figure~\ref{fig:QED}, and one can indeed see that they all correspond to
connected subdiagrams. This result was originally derived using a combinatoric
approach~\cite{Yennie:1961ad}, and has recently been rederived in a more
intuitive way using path integral methods and statistical physics 
techniques~\cite{Laenen:2008gt}, which pave the way for examining more
complicated cases of exponentiation e.g. multiparton scattering in QCD.
Note that exponentiation is a very powerful result: it tells us that we can
predict the structure of infrared divergences to all orders in perturbation 
theory. Successive powers of the coupling constant in the exponent sum up 
successive towers of IR singularities in the amplitude itself. Large logarithms
that are associated with IR singularities can also then be predicted to all
orders from soft exponentiation, which is how resummation works in practice.\\

The structure of soft exponentiation in QCD is more complicated due to the 
non-Abelian nature of the theory. Scattering amplitudes then have non-trivial
colour structure. Furthermore, one must draw a distinction between processes
in which only two coloured particles emerge from the hard interaction (e.g.
Drell-Yan production of vector bosons, deep-inelastic scattering, 
$e^+e^-\rightarrow q\bar{q}$), and processes in which many hard coloured 
particles scatter. In the latter case, the scattering amplitude becomes a 
vector in the space of possible colour flows at the hard interaction vertex, 
and the soft function has a matrix structure in this space. 
Soft exponentiation in the two-particle case was first studied 
in~\cite{Gatheral:1983cz}, where it was found that the corresponding webs 
consist of single, irreducible subdiagrams (that is, subdiagrams where
no gluon can be shrunk to the origin independently of any other gluon). The
nature of webs in multiparton scattering has been studied only very 
recently~\cite{Gardi:2010rn,Mitov:2010rp}, due to the involved colour 
structure remarked upon above. The results are in marked contrast to the 
two-line case, and are best illustrated here by example. \\

\begin{figure}
\begin{center}
\scalebox{0.8}{\includegraphics{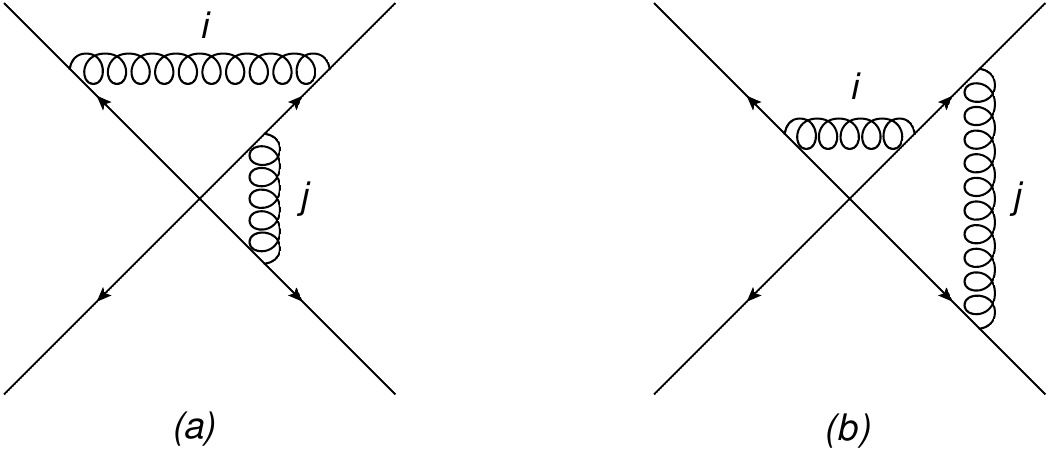}}
\caption{Diagrams contributing to the soft function at two loops, where
solid and wavy lines denote hard and soft particles respectively.}
\label{fig:2loop}
\end{center}
\end{figure}

Consider the pair of two-loop contributions to the soft function shown in 
figure~\ref{fig:2loop}, where each diagram $D$ has a kinematic part 
${\cal F}(D)$ and a colour factor $C(D)$. 
Although these diagrams are reducible (in the sense that one may shrink both
gluons separately to the origin in each case), one may show that both of them
contribute to the exponent of the soft function. Furthermore, each of them
has a modified colour factor $\tilde{C}(D)$ in the exponent, which turns out
to be a superposition of the usual colour factors $\{C(D)\}$ of both graphs.
The contribution to the exponent from the pair can be written 
\begin{equation}
  \left(\begin{array}{c}{\cal F}(a)\\{\cal F}(b)\end{array}\right)^T
  \left(\begin{array}{c}\tilde{C}(a)\\\tilde{C}(b)\end{array}\right)=
  \left(\begin{array}{c}{\cal F}(a) \\ {\cal F}(b)\end{array}\right)^T\frac{1}{2}\left(
  \begin{array}{rr}1& -1\\ -1& 1\end{array}\right)\left(\begin{array}{c}
      C(a)\\C(b)\end{array}\right).
\label{twoloop}
\end{equation}
We see that the pair of diagrams mixes under exponentiation, where this mixing
can be described by a matrix acting on the vectors of kinematic and colour 
factors.\\

This structure is found to be quite general. Higher-loop diagrams form 
closed sets, where each set has elements related by gluon permutations on the
external lines. It is argued in~\cite{Gardi:2010rn,Gardi:2011yz} that each set 
should be considered as a single web, and its contribution to the exponent of 
the soft function is given by the generic form
\begin{equation}
W=\sum_{D,D'\in W}{\cal F}(D)R_{DD'}C(D'),
\label{Rdef}
\end{equation}
where $R_{DD'}$ is a {\it web-mixing matrix}. These matrices consist of 
constant numbers, and encode a huge amount of physics, namely how colour
and kinematic information gets entangled in the soft limit to all orders in 
perturbation theory. An ongoing programme of work consists of classifying
general properties of these matrices, and interpreting the corresponding 
physics.\\

Some general properties are already known. For example, any row of any web 
mixing matrix must sum to zero. Also, all web mixing matrices are idempotent
(${\bf R}^2={\bf R}$), such that their eigenvalues can only be 1 or 
0~\cite{Gardi:2010rn,Gardi:2011wa}. A further conjecture involves a weighted
sum of column entries~\cite{Gardi:2011yz}. A pure combinatoric expression for
web mixing matrix elements has been given in~\cite{Gardi:2011wa}, and may
be related to order-preserving maps on partially ordered sets 
(posets)~\cite{Dukes}. The
latter are used in computer science applications, and thus progress in
understanding webs can be made with or without any field theory knowledge.\\

For massless external particles, it is conjectured that the exponent of the
soft function has a very simple form, involving colour and kinematic 
correlations between at most pairs of particles. This is known as the
{\it dipole formula}~\cite{Becher:2009cu}, and possible corrections (at three
loops and beyond) were further investigated in~\cite{Dixon:2009ur}. Recently,
new constraints were found to emerge from the high energy (Regge) limit of
scattering amplitudes~\cite{Bret:2011xm}, whose further implications were
examined in~\cite{Vernazza:2011aa}. Note that web mixing matrices should have
something to say about the dipole formula: both webs and the dipole formula are
concerned with the structure of the exponent of the soft function.

\section{IR singularities in gravity}
Infrared singularities in gravity were first examined 
in~\cite{Weinberg:1965nx}, and there has recently been a revival of interest
aimed at describing gravitational physics using the same language as is used
in non-abelian gauge theories~\cite{Akhoury:2011kq,Naculich:2011ry}. It has now
been firmly established that the exponent of the soft function in gravity is 
{\it one-loop exact} i.e. only one-loop diagrams occur, connecting all possible
pairs of external particles. The relationship between QED / QCD and gravity 
was further explored in~\cite{Miller:2012an}, using the radial coordinate space
picture of~\cite{Chien:2011wz}. This involves mapping the Wilson line operators
which describe soft photon or graviton emissions from Minkowski space to 
Euclidean AdS space, where they become point charges whose potential energy 
represents the cusp anomalous dimension. The general potential energy for a 
spin-$n$ Wilson line is found to be 
\begin{equation}
  \tilde{H}(\beta)=A_1\left(\frac{\sinh(n\beta)}{\sinh\beta}\right)
  +A_2\left(\frac{\cosh(n\beta)}{\sinh\beta}\right),
\label{potE}
\end{equation}
where $\beta$ is the radial distance in the AdS space (equivalent to the 
cusp angle in Minkowski space). Equation~(\ref{potE}) can be related to the 
known cusp anomalous dimensions in QED / QCD ($n=1$) and gravity ($n=2$). 
Furthermore, $n$ can be taken to be a continuous variable, and one thus sees 
that the soft limits of the two types of theory are related by a continuous 
deformation. This is an interesting novelty, that may have further 
implications.\\

Note that one-loop exactness of gravity implies that all IR singularities in
this theory are dipole-like. This begs the question: could the QCD dipole 
formula have a gravitational origin? This has been further explored 
in~\cite{Oxburgh:2012zr}, and the answer appears to be no. Nevertheless,
intriguing connections between QCD and gravity exist, whose investigation is
ongoing.

\end{document}